\documentclass[final,3p,times,authoryear]{elsarticle}

\makeatletter
\def\ps@pprintTitle{%
 \let\@oddhead\@empty
 \let\@evenhead\@empty
 \def\@oddfoot{\footnotesize\itshape
       \ifx\@journal\@empty Elsevier
       \else\@journal\fi\hfill\today}%
 \let\@evenfoot\@oddfoot}
\makeatother

\pdfoutput=1

\usepackage{amssymb}

\usepackage[displaymath, mathlines]{lineno}

\journal{Polar Science 10 (1), 11--23, 2016 (authors' version).
DOI: 10.1016/j.polar.2015.12.004.}

\newcommand{\bigfrac}[2]{\mbox{$\displaystyle\frac{#1}{#2}$}}
\newcommand{\pabl}[2]{\frac{\partial #1}{\partial #2}}
\newcommand{\bigpabl}[2]{\bigfrac{\partial #1}{\partial #2}}

\begin{document}

\begin{frontmatter}

\title{Comparison of thermodynamics solvers in the polythermal ice
sheet model SICOPOLIS}

\author[affil_ilts]{Ralf Greve}
\author[affil_ilts,affil_iaceth]{Heinz Blatter}

\address[affil_ilts]{Institute of Low Temperature Science,
Hokkaido University,
Kita-19, Nishi-8, Kita-ku, Sapporo 060-0819, Japan}
\address[affil_iaceth]{Institute for Atmospheric and Climate Science,
ETH Zurich,
Universit\"atstrasse 16, CH-8092 Zurich, Switzerland \\[1ex]
Corresponding author: Ralf Greve (greve@lowtem.hokudai.ac.jp)}

\begin{abstract}
In order to model the thermal structure of polythermal ice sheets
accurately, energy-conserving schemes and correct tracking of the
cold--temperate transition surface (CTS) are necessary. We compare four
different thermodynamics solvers in the ice sheet model SICOPOLIS. Two
exist already, namely a two-layer polythermal scheme (POLY) and a
single-phase cold-ice scheme (COLD), while the other two are
newly-implemented, one-layer enthalpy schemes, namely a conventional
scheme (ENTC) and a melting-CTS scheme (ENTM). The comparison uses
scenarios of the EISMINT Phase~2 Simplified Geometry Experiments
\citep[][J.~Glaciol.\ 46, 227--238]{Eismint2000}. The POLY scheme is
used as a reference against which the performance of the other schemes
is tested. Both the COLD scheme and the ENTC scheme fail to produce a
continuous temperature gradient across the CTS, which is explicitly
enforced by the ENTM scheme. ENTM is more precise than ENTC
for determining the position of the CTS, while the performance of
both schemes is good for the temperature/water-content profiles in
the entire ice column. Therefore, the one-layer enthalpy schemes ENTC
and ENTM are viable, easier implementable alternatives to the POLY
scheme with its need to handle two different numerical domains for cold
and temperate ice.
\end{abstract}


\end{frontmatter}

\nolinenumbers

\section{Introduction}
\label{sec:intro}

Many glaciers and ice sheets are polythermal with disjoint cold and
temperate domains, separated by the cold--temperate transition surface
(CTS) \citep{Blatter1991}. Both the Greenland and Antarctic ice sheets
are Canadian-type polythermal, that is, they are mainly cold, except for
distributed temperate layers at the base where strain heating is
largest and where there is a geothermal contribution. It is thus
important to model the thermodynamics of ice sheets
correctly by distinguishing both domains and accounting for the
transition conditions between them.

Various methods allow one to model the thermodynamic conditions in ice
sheets. Thus far, SICOPOLIS (SImulation COde for POLythermal Ice Sheets;
e.g., \citeauthor{Greve1997b}, \citeyear{Greve1997b};
\citeauthor{sato_greve_2012_annglac},
\citeyear{sato_greve_2012_annglac};
\citeauthor{greve_herzfeld_2013_annglac},
\citeyear{greve_herzfeld_2013_annglac}; URL www.sicopolis.net)
is the only three-dimensional ice sheet model that employs the
\emph{polythermal two-layer scheme}. In this method, the temperature and
water-content fields in the two domains, cold and temperate ice, are
computed on separate numerical domains, and the transition conditions at
the CTS are used to track its position.

In most older ice sheet models \citep[e.g.,][]{huybrechts_1990a,
calov_hutter_1996, payne_dongelmans_1997, ritz_etal_1997}, the
\emph{cold-ice method} was applied by resetting any computed
temperatures that exceed the local pressure melting point to the local
pressure melting point. While very simple, this means that energy is
lost, and the water content in the temperate layer as well as the
transition conditions at the CTS are ignored. The cold-ice method has,
however, always been available in SICOPOLIS as an alternative to the
polythermal two-layer method.

\citet{AschwandenBueler2012} introduced a new, enthalpy-based approach
for ice sheet thermodynamics. In this method, the thermodynamic fields
of temperature in cold ice and water content in temperate ice are
replaced by one common thermodynamic field, enthalpy\footnote{Owing to
the incompressibility of ice, the enthalpy is identical to the internal
energy.}, for both domains, and only one common field equation must be
solved. However, the Stefan-type energy- and mass-flux matching
conditions at the CTS, which are important for determining its position
\citep{Greve1997}, are not included explicitly in the formulation of the
enthalpy scheme by \citet{AschwandenBueler2012}. Following the
terminology of \citet{Blatter&Greve2015}, we refer to it as the
\emph{conventional one-layer enthalpy scheme}. This scheme has already
been used in a number of ice sheet and glacier models
\citep{Brinkerhoff&Johnson2013, Golledge&al2013, Seroussi&al2013,
Wilson&Flowers2013, Gilbert&al2014, Kleiner&al2015}.

Two different conditions of the CTS must be distinguished. For melting
conditions, cold ice flows across the CTS into the temperate layer,
where water starts to accumulate due to strain heating along
trajectories. The opposite situation, freezing conditions, occurs
further downstream, where temperate ice flows across the CTS into the
cold domain and the accumulated water content freezes out, releasing
latent heat. For melting conditions, the temperature gradient and the
water content are continuous across the CTS, while for freezing
conditions, discontinuities of these quantities occur \citep{Greve1997}.
Since the CTS tends to be rather steep near the terminus, only a small
area of the CTS is freezing, and therefore, in results of ice sheet
models, freezing conditions usually only occur at very few isolated grid
points \citep{Greve1997b}.

\citet{Kleiner&al2015} tested the implementation of the conventional
enthalpy scheme for a Canadian-type parallel-sided slab in one
finite-dif\-fe\-rence and two finite-element ice sheet models
(TIM-FD$^3$, ISSM, COMice). \citet{Blatter&Greve2015} compared the
performance of four different versions of the enthalpy scheme for a
parallel-sided slab with a custom-designed finite-difference program.
Besides the conventional enthalpy scheme, they considered the
\emph{two-layer front-tracking enthalpy scheme} (an enthalpy version of
the polythermal two-layer scheme mentioned above), the \emph{one-layer
melting-CTS enthalpy scheme} and the \emph{one-layer freezing-CTS
enthalpy scheme}. In the two latter schemes, explicit tracking of the
melting or freezing CTS, based on the respective transition conditions
at the CTS, has been added to the conventional enthalpy scheme. An
important finding of these works was that the conventional one-layer
enthalpy scheme can produce correct solutions for melting conditions at
the CTS, provided that the numerical handling of the discontinuity of
the enthalpy diffusivity across the CTS is done carefully. However,
especially for finite-difference techniques, \citet{Blatter&Greve2015}
concluded that it is safer to use the one-layer melting-CTS enthalpy
scheme, which enforces the transition conditions explicitly. For
freezing conditions, the conventional one-layer enthalpy scheme fails
because it cannot handle the associated discontinuities of the
thermodynamic fields, and it is thus imperative to enforce the
transition conditions at the CTS explicitly, as it is done in the
one-layer freezing-CTS enthalpy scheme.

For this study, in addition to the previously existing polythermal
two-layer and cold-ice schemes, we have implemented the conventional
one-layer enthalpy scheme and the one-layer melting-CTS enthalpy scheme
in the SICOPOLIS model. For the reason given above, freezing conditions
are not considered here. We attempt to test and verify these four schemes in
SICOPOLIS, and in particular to test how the various schemes handle the
melting CTS between cold and temperate ice for Canadian-type
polythermal situations in ice sheets. Based on the results of
\citet{Blatter&Greve2015}, we consider the polythermal two-layer scheme
to be the most reliable method and thus use its results as benchmark
solutions. In Sections~\ref{sec:thermodynamics} and
\ref{sec:thermodyn_solvers} we give an overview of the theory of
ice-sheet thermodynamics and describe the implementation of the various
schemes in SICOPOLIS. Section~\ref{sec:setup} gives the set-up of the
scenarios derived from the suite of EISMINT (European Ice Sheet Modeling
INiTiative) Phase~2 Simplified Geometry Experiments \citep{Eismint2000}
used for this study. In Section~\ref{sec:results} we discuss the
results, focusing on the simulated thickness of the temperate ice layer.
Section~\ref{sec:discussion} concludes the paper.

\section{Outline of ice-sheet thermodynamics}
\label{sec:thermodynamics}

\subsection{Standard polythermal thermodynamics}
\label{ssec:thermodynamics_1}

The standard description of the thermodynamics of polythermal
ice mas\-ses, for which we follow largely \citet{Greve1997},
is based on the fields of absolute temperature $T$ in cold ice
and water content $W$ in temperate ice. The evolution equation
for temperature in cold ice is given by
\begin{equation}
  \label{eq:temperature_c}
  \frac{\partial T}{\partial t} + \mathbf{v}\cdot\mathrm{grad}\,T
  = \frac{1}{\rho{}c}\,
    \pabl{}{z}\left(\kappa\pabl{T}{z}\right)
    + \frac{Q}{\rho{}c}\,,
\end{equation}
where $t$ denotes time, $z$ the vertical spatial coordinate,
$\mathbf{v}$ the three-di\-men\-sio\-nal velocity vector,
$\rho=910\,\mathrm{kg\,m^{-3}}$ the ice density, $\kappa$ the
temperature-dependent heat conductivity of cold ice and $c$ the
temperature-dependent heat capacity of cold ice. Also,
$Q=\mathrm{tr}(\mathbf{t}\cdot\mathbf{D})$ is the volumetric strain
heating, where $\mathbf{t}$ is the Cauchy stress tensor, $\mathbf{D}$
the strain-rate tensor, the middle dot $(\cdot)$ denotes tensor
contraction and tr the trace of a tensor. Horizontal diffusion terms
have been neglected, which can be justified by scaling arguments making
use of the shallowness of ice sheets \citep[e.g.][]{GreveBlatter2009}.

Similar to Eq.~(\ref{eq:temperature_c}), the evolution equation for
water content in temperate ice reads
\begin{equation}\begin{array}[b]{l}
  \label{eq:water_content_t}
  \bigpabl{W}{t} + \mathbf{v}\cdot\mathrm{grad}\,W
  = \bigfrac{1}{\rho}\,
    \bigpabl{}{z}\left(\nu\bigpabl{W}{z}\right)
    + \bigfrac{Q}{\rho{}L}
  \\[3ex]\hspace*{3em}
    -\bigfrac{c}{L}
     \left( \bigpabl{T_\mathrm{m}}{t}
            + \mathbf{v}\cdot\mathrm{grad}\,T_\mathrm{m} \right)
    + \bigfrac{1}{\rho{}L}\,
      \bigpabl{}{z}\left(\kappa\bigpabl{T_\mathrm{m}}{z}\right)\,,
\end{array}\end{equation}
where $\nu$ is the water diffusivity in temperate ice
(assumed to be constant) and
$L=3.35{}\times{}10^5\,\mathrm{J\,kg^{-1}}$ the latent heat
of fusion. The very small terms in the second line of the equation
arise from the fact that the temperature in temperate ice is not
constant, but equal to the melting temperature $T_\mathrm{m}$
that depends on the local pressure $p$,
\begin{equation}
  T_\mathrm{m}(p) = T_0 - \beta p\,,
  \label{eq:temp-melt}
\end{equation} 
where $T_0=273.15\,\mathrm{K}$ is the reference temperature
and $\beta=9.8\times{}10^{-8}\,\mathrm{K\,Pa^{-1}}$ the
Clausius-Clapeyron constant for air-saturated glacier ice
\citep{Hooke2005}. As in Eq.~(\ref{eq:temperature_c}),
horizontal diffusion terms have been neglected in the evolution
equation (\ref{eq:water_content_t}).

As already mentioned in Sect.~\ref{sec:intro}, for melting conditions
at the CTS, the temperature gradient and the water content must be
continuous across the CTS \citep{Greve1997}. If we mark values at the
cold side of the CTS by plus (+) superscripts, values at the
temperate side by minus ($-$) superscripts, and denote the normal
unit vector pointing into the cold side by $\mathbf{n}$, this
reads
\begin{equation}
  \mathrm{grad}\,T^+\cdot\mathbf{n}
  = \mathrm{grad}\,T_\mathrm{m}^-\cdot\mathbf{n}
  \label{eq:gradT-CTS}
\end{equation}
and
\begin{equation}
  W^+ = W^- = 0\,.
  \label{eq:water-CTS}
\end{equation}

For freezing conditions, the situation is more complicated in that
the temperature gradient and the water content are in general
discontinuous across the CTS \citep{Blatter1991, Greve1997}.
However, as already mentioned in Sect.~\ref{sec:intro}, this
situation usually occurs only on very small parts of the CTS,
and therefore we do not consider freezing conditions in this study.

At the ice surface, we prescribe the surface temperature as a
Dirichlet-type boundary condition. At the ice base, three different
cases must be distinguished. For a cold base, the geothermal heat
flux determines the normal derivative of the temperature
(Neumann-type boundary condition). For a temperate base with no
temperate ice layer above, the temperature is equal to the pressure
melting point. For a temperate base with a temperate ice layer above,
in addition to that, a boundary condition for the basal water content
is required (unless the water diffusivity $\nu$ is neglected).
In order to influence the solution as little as possible, 
we choose the Neumann-type zero-flux condition
$\partial{}W/\partial{}z=0$.

In order to prevent the water content in temperate ice from rising
to unreasonable levels due to the accumulated strain heating,
a drainage model is required. We have kept the simple formulation
previously implemented in SICOPOLIS, which assumes that any water
exceeding the prescribed threshold of $W_\mathrm{max}=0.01$ ($=1\%$)
is drained instantaneously. The transport mechanism to the ice base
remains unmodeled; however, the amount of drained water is added
to the computed basal melt rate, and thus accounted for in the
computation of the vertical velocity and the evolution of the ice
thickness.

\subsection{Enthalpy method}
\label{ssec:thermodynamics_2}

As an alternative to the polythermal thermodynamics outlined in
Sect.~\ref{ssec:thermodynamics_1}, \citet{AschwandenBueler2012}
introduced the enthalpy method. Its strength is that it combines the
temperature and water content into a single thermodynamic field, the
specific enthalpy, for which a single field equation holds.

The specific enthalpy $h$ of ice at absolute temperature $T$ and water
content $W$ is given by
\begin{equation}
  \label{eq:enthalpy-ice}
  h(T,W) 
  = \int\limits_{T_0}^T c(\tilde{T})\,\mathrm{d}\tilde{T} + LW\,.
\end{equation} 
Inversely, the temperature and water content are not unique functions
of the enthalpy because of the dependence of the melting temperature
on the pressure (Eq.~(\ref{eq:temp-melt})).
We denote the inverse of Eq.~(\ref{eq:enthalpy-ice})
for zero water content ($W=0$) by $T(h)$ and the enthalpy of ice
at the melting point for zero water content by $h_\mathrm{m}$,
\begin{equation}
  \label{eq:enthalpy-ice-m}
  h_\mathrm{m}(p)
  = h(T_\mathrm{m}(p),W\!=\!0)
  = \int\limits_{T_0}^{T_\mathrm{m}(p)}
       c(\tilde{T})\,\mathrm{d}\tilde{T}\,.
\end{equation} 
The temperature and water content can then be obtained by
\begin{eqnarray}
  \label{eq:temp-from-enthalpy}
  T(h,p) &=& \left\{ \begin{array}{ll}
         T(h)\,,           & h < h_\mathrm{m}(p)\,, \\
         T_\mathrm{m}(p)\,, & h \ge h_\mathrm{m}(p)\,,
         \end{array}  \right. 
  \\[1ex]
  \label{eq:omega-from-enthalpy}
  W(h,p) &=& \left\{ \begin{array}{ll}
         0\,,                        & h < h_\mathrm{m}(p)\,, \\
         (h-h_\mathrm{m}(p))/L\,, & h \ge h_\mathrm{m}(p)\,.
         \end{array}\right.
\end{eqnarray} 
The balance equation for enthalpy reads
\begin{equation}
  \label{eq:enthalpy} 
  \frac{\partial h}{\partial t} + \mathbf{v}\cdot\mathrm{grad}\,h
   = \frac{\partial}{\partial z}\,  
  \left( k_\mathrm{c,t}\,  \frac{\partial h}{\partial z} \right)
  + \frac{Q}{\rho}\,,
\end{equation} 
where 
\begin{equation}
  \label{eq:enthalpy-diffusivity}
  k_\mathrm{c,t} =
  \left\{\begin{array}{ll}
    k_\mathrm{c} = \mbox{$\displaystyle\frac{\kappa}{\rho c}$}\,,
       \; & h < h_\mathrm{m}(p)\,,
    \\[3ex]
    k_\mathrm{t} = \mbox{$\displaystyle\frac{\nu}{\rho}$}\,,
       \; & h \ge h_\mathrm{m}(p)
  \end{array}\right.
\end{equation}
are the enthalpy diffusivities in cold (subscript c) and temperate
(subscript t) ice, respectively. Like in Eqs.~(\ref{eq:temperature_c})
and (\ref{eq:water_content_t}), horizontal diffusion terms have been
neglected in Eq.~(\ref{eq:enthalpy}), while vertical diffusion is
present in both cold and temperate ice.

The transition conditions for a melting CTS,
Eqs.~(\ref{eq:gradT-CTS}) and (\ref{eq:water-CTS}), transform
into physically equivalent enthalpy conditions.
Inserting Eq.~(\ref{eq:water-CTS}) in Eq.~(\ref{eq:enthalpy-ice})
and considering that the temperature on both sides of the CTS
is equal to the pressure melting point $T_\mathrm{m}$ yields the
continuity of the enthalpy,
\begin{equation}
  h^+ = h^- = h_\mathrm{m}\,.
  \label{eq:enth-CTS}
\end{equation}
On the cold side of the CTS, $W=0$ holds everywhere, so that
differentiating Eq.~(\ref{eq:enthalpy-ice}) yields
$\mathrm{grad}\,h^+=c(T_\mathrm{m})\,\mathrm{grad}\,T^+$. On the
temperate side, differentiating Eq.~(\ref{eq:enthalpy-ice-m}) yields
$\mathrm{grad}\,h_\mathrm{m}^-
=c(T_\mathrm{m})\,\mathrm{grad}\,T_\mathrm{m}^-$. Hence,
the enthalpy form of Eq.~(\ref{eq:gradT-CTS}) is
\begin{equation}
  \mathrm{grad}\,h^+\cdot\mathbf{n}
    = \mathrm{grad}\,h_\mathrm{m}^-\cdot\mathbf{n}\,,
  \label{eq:gradh-CTS}
\end{equation}
which expresses the continuity of the sensible heat flux. Note that
the enthalpy gradient across the CTS is in general discontinuous 
($\mathrm{grad}\,h^+\cdot\mathbf{n}>\mathrm{grad}\,h^-\cdot\mathbf{n}$)
because the enthalpy in the temperate ice layer becomes larger than
$h_\mathrm{m}$ away from the CTS due to the increasing water content.

For freezing conditions, in general the enthalpy and the sensible
heat flux are discontinuous ($h^+<h^-$, 
$\mathrm{grad}\,h^+\cdot\mathbf{n}
<\mathrm{grad}\,h_\mathrm{m}^-\cdot\mathbf{n}$).
The exact form of the transition conditions shall not be given here.

By employing Eq.~(\ref{eq:enthalpy-ice}), the boundary conditions for
the ice surface and base, as well as the simple drainage model, given in
the last two paragraphs of Sect.~\ref{ssec:thermodynamics_1} translate
readily to the enthalpy formulation. However, for a temperate base with
a temperate ice layer above, instead of the zero-flux condition
for the water content $\partial{}W/\partial{}z=0$ we use the zero
flux condition for the enthalpy $\partial{}h/\partial{}z=0$. Since the
temperature in temperate ice is not constant (Eq.~(\ref{eq:temp-melt})),
these conditions are not equivalent, but the difference is small and both
conditions are ad-hoc anyway, so that this is acceptable for the sake of
simplicity.

\section{Thermodynamics solvers in SICOPOLIS}
\label{sec:thermodyn_solvers}

Previous versions of SICOPOLIS (3.1 and older) contained two different
options for dealing with ice-sheet thermodynamics. These are the
polythermal two-layer scheme and the cold-ice scheme. Here, we added two
more options based on the enthalpy method and the schemes described by
\citet{Blatter&Greve2015}, namely the conventional one-layer enthalpy
scheme and the one-layer melting-CTS enthalpy scheme. Freezing
conditions at the CTS are not considered.

\citet{Blatter&Greve2015} elaborated the schemes only for a
one-di\-men\-sio\-nal problem (parallel-sided slab). Here, we extended
them to three dimensions. SICOPOLIS uses terrain-following (``sigma'')
coordinates, and the transformation to these coordinates produces extra
terms \citep[e.g.,][Sect.~5.7.1]{GreveBlatter2009}. However, since
horizontal diffusion has been neglected in Eq.~(\ref{eq:enthalpy}), even
in the transformed coordinates no mixing of horizontal and vertical
derivatives of the enthalpy $h$ occurs. This allows employing an
implicit discretization scheme for the vertical derivatives and an
explicit scheme for the horizontal derivatives in the same way as for
the temperature and water-content equations (\ref{eq:temperature_c}) and
(\ref{eq:water_content_t}) \citep{Greve1997b}. Thus, the numerical
solution for each vertical profile is similar to the solution of the
one-dimensional problem by an implicit scheme, and the horizontal
advection terms of the three-dimensional equation play the role of
additional, explicitly discretized source terms.

\subsection{Polythermal two-layer scheme (POLY)}
\label{ssec:thermodyn_solvers_1}

The polythermal two-layer scheme (scheme code: POLY) is the most
sophisticated, but also the most complex method to simulate
polythermal ice masses. It splits the computational domain
numerically into two distinct layers of cold and temperate ice, and
solves the evolution equations for temperature in cold ice
(Eq.~(\ref{eq:temperature_c})) and water content in temperate ice
(Eq.~(\ref{eq:water_content_t})) on two different grids
\citep{Blatter1991, Greve1997, Pettersson2007}. In this method, the CTS
is fixed with the lower and upper boundaries of the cold and temperate
domains, respectively, and thus can be tracked very precisely with the
transition condition at the CTS.

The method works for both melting and freezing conditions at the CTS;
in particular, it can cope with the discontinuities of the
temperature gradient and the water content that accompany
freezing conditions. It is implemented in SICOPOLIS as an iterative
trial-and-error procedure. For each time step and each ice column for
which an internal CTS is detected, the temperature problem for the upper
(cold-ice) region and the water-content problem for the lower
(temperate-ice) region are solved repeatedly with different test
positions of the CTS until all boundary and transition conditions are
fulfilled. This is discussed in more detail by \citet[][last paragraph
of section~2]{Greve1997b}. For the simulations discussed in this study,
we disregard freezing conditions and apply the POLY
scheme only for assumed melting conditions at the CTS. The ability to
turn off the freezing conditions is a newly-implemented option for the
POLY scheme.

\subsection{Cold-ice scheme (COLD)}
\label{ssec:thermodyn_solvers_2}

The cold-ice scheme (scheme code: COLD) is the most simple, yet
physically incorrect way to deal with polythermal conditions in ice
sheets. SI\-CO\-PO\-LIS applies the COLD scheme by solving the temperature
equation (\ref{eq:temperature_c}) for the entire domain and resetting
any temperatures $T$ exceeding the local pressure melting point
$T_\mathrm{m}$ (Eq.~(\ref{eq:temp-melt})) to $T=T_\mathrm{m}$. The
excess heat of the temperate layer becomes lost, so that the scheme does
not conserve energy.

\subsection{Conventional one-layer enthalpy scheme (ENTC)}
\label{ssec:thermodyn_solvers_3}

The conventional one-layer enthalpy scheme (scheme code: ENTC)
corresponds to the enthalpy method by \citet{AschwandenBueler2012}. The
enthalpy equation (\ref{eq:enthalpy}) is solved for the entire
polythermal domain on a single numerical grid, and the CTS is
diagnostically determined as the uppermost grid point of the temperate
layer, that is, the uppermost grid point for which $h\ge{}h_\mathrm{m}$
holds. The transition conditions at the CTS, Eqs.~(\ref{eq:enth-CTS})
and (\ref{eq:gradh-CTS}) for melting conditions or their more
complicated counterparts for freezing conditions, are not enforced
explicitly.

\subsection{One-layer melting-CTS enthalpy scheme (ENTM)}
\label{ssec:thermodyn_solvers_4}

The one-layer melting-CTS enthalpy scheme (scheme code: ENTM) by
\citet{Blatter&Greve2015} differs from the conventional one-layer
enthalpy (ENTC) scheme in that the continuity condition 
(\ref{eq:gradh-CTS}) at a melting CTS is enforced
explicitly. This is achieved for each vertical column possessing an
internal CTS (that is, a non-vanishing thickness of temperate ice at the
base) in two steps. In a predictor step, the enthalpy profile is
computed for the entire vertical column like in the ENTC
scheme. The CTS is then positioned at the uppermost
point in the temperate layer. In a corrector step, the enthalpy profile
in the upper, cold layer only is recomputed by using condition
(\ref{eq:gradh-CTS}) as lower boundary condition. The updated enthalpy
profile then consists of the predictor step for the temperate layer and
the corrector step for the cold layer. For the technical details of the
scheme see \citet{Blatter&Greve2015}.

The implementation of both the ENTC and the ENTM scheme in
SI\-CO\-PO\-LIS has the same topological restriction as the POLY scheme.
It only allows for Canadian-type polythermal conditions, that is, only
one CTS can exist in each column of ice, and the cold layer is always
situated above the temperate layer. However, this is not a fundamental
restriction for the enthalpy schemes. In principle, they also allow
dealing with more complex and/or changing topologies such as temperate
ice at the surface, prescribed water content instead of prescribed
temperature as a boundary condition at the surface, bubbles of
temperate ice within cold ice etc.

\section{Experiment set-up}
\label{sec:setup}

We tested the four different schemes described in
Sect.~\ref{sec:thermodyn_solvers} with modified versions
of the EISMINT Phase~2 Simplified
Geometry Experiments~A and H as defined in \citet{Eismint2000}.
The computational domain in
the horizontal plane is a square of 1500 by 1500~km, spanned by the
Cartesian coordinates $x=0\ldots{}1500\,\mathrm{km}$,
$y=0\ldots{}1500\,\mathrm{km}$. The numerical values of the dynamical
and thermodynamical parameters are given in
\citet{Eismint2000} (see also Sect.~\ref{sec:thermodynamics}). The
experiments were performed using SICOPOLIS version 3.2-dev (revision
471) with three different combinations of horizontal grid resolution
$\Delta{}x$ ($=\Delta{}y$) and time step $\Delta{}t$, namely
$(\Delta{}x,\,\Delta{}t)=(10\,\mathrm{km},\,2\,\mathrm{a})$,
$(5\,\mathrm{km},\,2\,\mathrm{a})$ and
$(10\,\mathrm{km},\,20\,\mathrm{a})$. For the 10-km resolution, this
leads to $151\times{}151$ grid points (indices $i=0\ldots{}150$,
$j=0\ldots{}150$) in the horizontal domain, while for the 5-km
resolution the domain is covered by $301\times{}301$ grid points.
The standard vertical
resolution is 81 grid points in the upper domain (terrain-following
vertical coordinate $\zeta_\mathrm{c}=0\ldots{}1$, index
$k_\mathrm{c}=0\ldots{}80$) and, for the POLY scheme,
11 grid points in the lower domain (terrain-following vertical
coordinate $\zeta_\mathrm{t}=0\ldots{}1$, index
$k_\mathrm{t}=0\ldots{}10$). For the other three schemes, the lower
domain is not used. For the $(10\,\mathrm{km},\,2\,\mathrm{a})$
combination, we also consider a five times higher vertical resolution
(abbreviated as ``hvr'') with 401 grid points in the upper and 51 grid
points in the lower domain.

Both the heat capacity $c$ and the heat conductivity $\kappa$ depend
significantly on temperature \citep[e.g.][]{GreveBlatter2009}. The
implementation of all thermodynamics schemes in SICOPOLIS accounts for
that possibility by allowing both quantities to vary in space and time.
This is particularly relevant for discretizing the vertical diffusion
terms in Eqs.~(\ref{eq:temperature_c}), (\ref{eq:water_content_t})
and (\ref{eq:enthalpy}) that contain $\kappa$ within 
$\partial{}/\partial{}z$ terms.
However, in the EISMINT set-up used here the values
are taken as constants, $c=2009\;\mathrm{J\,kg^{-1}K^{-1}}$ and
$\kappa=2.1\;\mathrm{W\,m^{-1}K^{-1}}$ \citep{Eismint2000}. The
diffusive flux in the temperate layer is likely very small for
small water contents; however, virtually no information on the value of
the water diffusivity $\nu$ is available. In SICOPOLIS, a value of
$\nu=10^{-6}\,\mathrm{kg\,m^{-1}\,s^{-1}}$ is implemented for reasons of
numerical stability rather than for physical reasons. This leads to
$k_\mathrm{t}=\nu/\rho\approx{}1.1\times{}10^{-9}\,\mathrm{m^{2}\,s^{-1}}$,
while $k_\mathrm{c}=\kappa/(\rho{}c)
\approx{}1.1\times{}10^{-6}\,\mathrm{m^{2}\,s^{-1}}$, thus a three orders
of magnitude difference.

\subsection{EISMINT experiment A1}
\label{ssec:setup1}

The experiment is started with no-ice conditions and runs for 200~ka to reach 
a steady state. The ice sheet rests on a flat, rigid base (no isostatic
adjustment). The boundary conditions are circularly symmetric with 
respect to the center of the domain. The prescribed surface temperature
increases linearly with distance from the center, with a minimum value
of 238.15~K at the center (summit) and a horizontal gradient of 
$1.67\times{}10^{-2}\;\mathrm{K\,km^{-1}}$. Similarly,
the surface mass balance is prescribed by a piece-wise linear function 
with a maximum accumulation rate of $0.5\;\mathrm{m\,a^{-1}}$ in the
interior of the ice sheet (from the center to 400~km away from the center),
and a horizontal gradient of $-10^{-2}\;\mathrm{m\,a^{-1}\,km^{-1}}$
beyond that, which leads to an equilibrium line 450~km from the center
and an ablation zone further out. With these assumptions, the evolving
ice sheet thickness does not feed back on the surface mass balance.
At the ice base, no-slip conditions are assumed everywhere, and the
geothermal heat flux is $42\;\mathrm{mW\,m^{-2}}$.

There are two differences compared to the original EISMINT experiment A.
The first one is the higher horizontal resolution; we use 10 and 5\,km
(see above) compared to the original 25\,km. The second one is that,
following \citet{Lliboutry&Duval1985}, we employ a water-content-dependent
rate factor $A_\mathrm{t}$ for temperate ice,
\begin{equation}
  A_\mathrm{t}(W) = A_{0^\circ\mathrm{C}}\times(1 + 1.8125\,W[\%])\,,
  \label{eq:rate-factor-temp1}
\end{equation}
where
$A_{0^\circ\mathrm{C}}=4.529\times{}10^{-24}\,\mathrm{s^{-1}\,Pa^{-3}}$
is the rate factor for $0^\circ\mathrm{C}$ (relative to pressure melting)
that results from the original EISMINT set-up.

\subsection{EISMINT experiment A2}
\label{ssec:setup2}

The set-up of experiment~A2 is essentially the same as that of
experiment~A1; however, the water-content-dependent rate factor
(\ref{eq:rate-factor-temp1}) is not used. Instead, the original EISMINT
set-up is used, which only considers the dependence of the rate factor
on temperature. Physically, this means that the lubrication effect
(through enhanced shear) of the near-basal temperate ice due to its
water content is ignored.

\subsection{EISMINT experiment H1}
\label{ssec:setup3}

Experiment H1 replaces the no-slip condition of A1 by basal sliding.
Otherwise, the set-up is the same. The sliding law is most simple and
relates the sliding velocity linearly to the basal drag. In the original
experiment~H, the sliding parameter has the value
$10^{-3}\;\mathrm{m\,a^{-1}\,Pa^{-1}}$. However, this does not produce
any basal temperate layers except for the COLD scheme due to the strong,
downward advection of cold surface ice. Since we want to focus on the
temperate ice, we decided to use a modified set-up, with the sliding
parameter reduced by a factor 10 to the value
$C_\mathrm{b}^0=10^{-4}\;\mathrm{m\,a^{-1}\,Pa^{-1}}$. 

Further, in the EISMINT set-up, basal sliding is assumed to occur
wherever the ice base is at the pressure melting point, while for
a cold base no-slip conditions prevail. However, when used with
the shallow ice approximation (that is employed by SICOPOLIS),
this binary switch is known to produce a singularity of the vertical
velocity field that prohibits proper convergence of the numerical
solution \citep{Bueler&Brown2009}. In order to circumvent this
problem, we regularize the transition by allowing for sub-melt
sliding in the form by \citet{Greve2005a} (originally proposed
by \citet{Hindmarsh&LeMeur2001})\footnote{An alternative way of
regularizing the singularity at the spatial onset of basal sliding
is to replace the shallow-ice stress balance by a higher-order one
including membrane stresses, such as the first order approximation
\citep{Blatter1995} or a shallow-ice--shelfy-stream hybrid
\citep{Bueler&Brown2009, pollard_deconto_2012b}.}.
The sliding parameter $C_\mathrm{b}$ depends on the basal temperature
$T_\mathrm{b}'$ (in ${}^\circ\mathrm{C}$, relative to pressure melting)
via
\begin{equation}
  C_\mathrm{b}
  = C_\mathrm{b}^0\,\mathrm{exp}(T_\mathrm{b}'/\gamma)\,,
  \label{eq:sub-melt-sliding}
\end{equation}
where $\gamma=1^\circ\mathrm{C}$ is the sub-melt-sliding parameter.
Equation (\ref{eq:sub-melt-sliding}) has the effect that, rather than
stopping abruptly, basal sliding decays exponentially as the basal
temperature falls below the pressure melting point.

\section{Results}
\label{sec:results}

\begin{figure}[htb]
\begin{center}
\includegraphics[width=90mm]{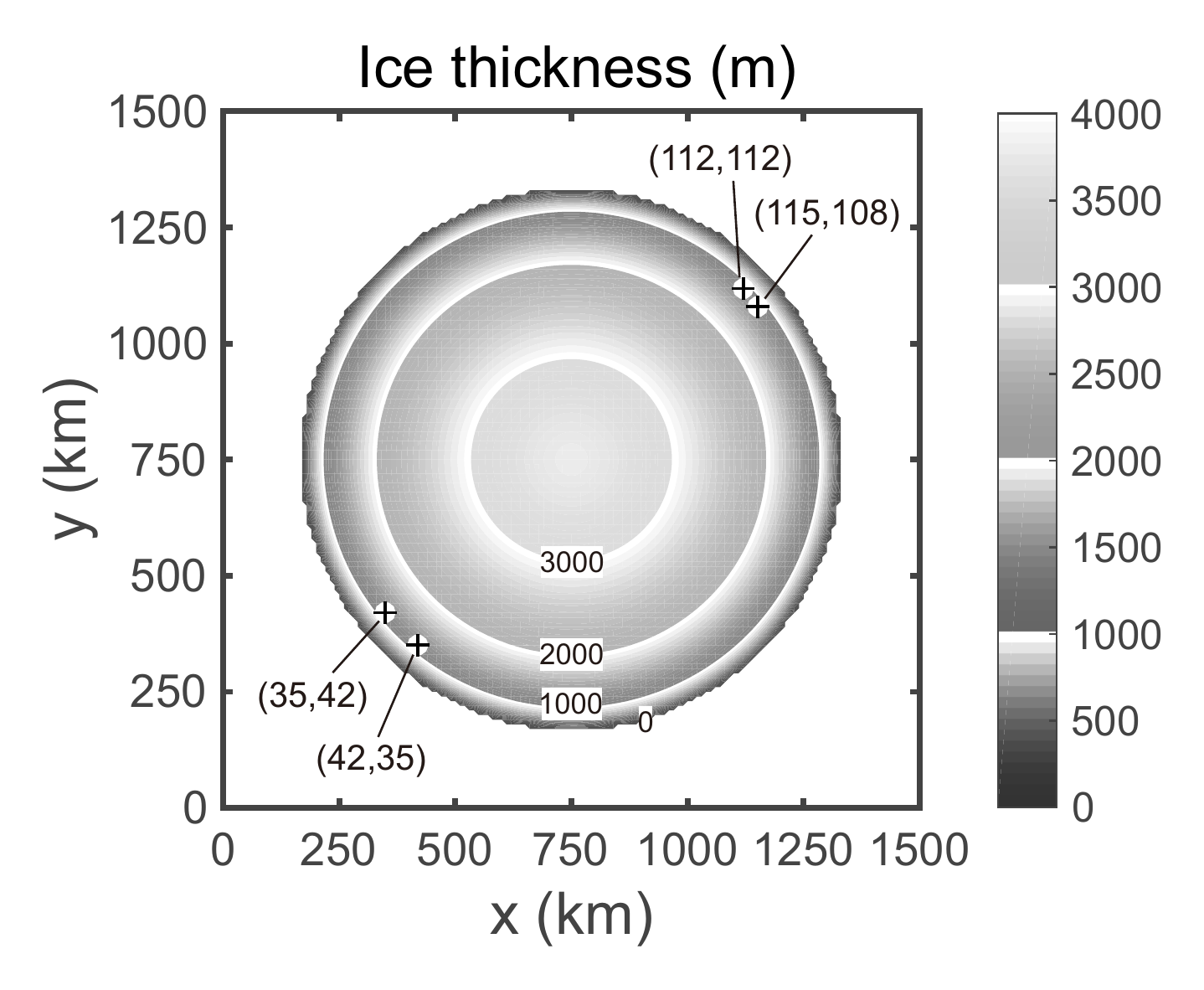}
\caption{EISMINT experiment~A1: 
Simulated ice thickness (=~surface topography) after 200\,ka model time.
Polythermal two-layer (POLY) scheme,
grid resolution $\Delta{}x=10\,\mathrm{km}$,
time step $\Delta{}t=2\,\mathrm{a}$.
The positions of the four profiles shown in
Fig.~\ref{fig:sico_plots_profiles} are marked by their respective
grid indices $(i,j)$.
Within grid resolution, they are at the same distance from the center
[(35,42), (42,35) and (115,108) 518.6\,km away,
(112,112) 523.3\,km away].}
\label{fig:expA_10km_20a_surftopo}
\end{center}
\end{figure}

For experiment~A1, run with the polythermal two-layer (POLY) scheme and
the parameters $\Delta{}x=10\,\mathrm{km}$, $\Delta{}t=2\,\mathrm{a}$,
Fig.~\ref{fig:expA_10km_20a_surftopo} depicts the simulated
steady-state ice thickness (=~surface topography) at the end of the
simulation ($t=200\,\mathrm{ka}$). The ice sheet reaches a volume of
$2.109\times{}10^6\,\mathrm{km}^3$, covers an area of
$1.046\times{}10^6\,\mathrm{km}^2$ and reaches a maximum thickness of
$3.687\,\mathrm{km}$. The melt fraction (fraction of basal ice at the
pressure melting point) is 67.5\%.

\begin{figure}[htb]
\begin{center}
\includegraphics[width=100mm]{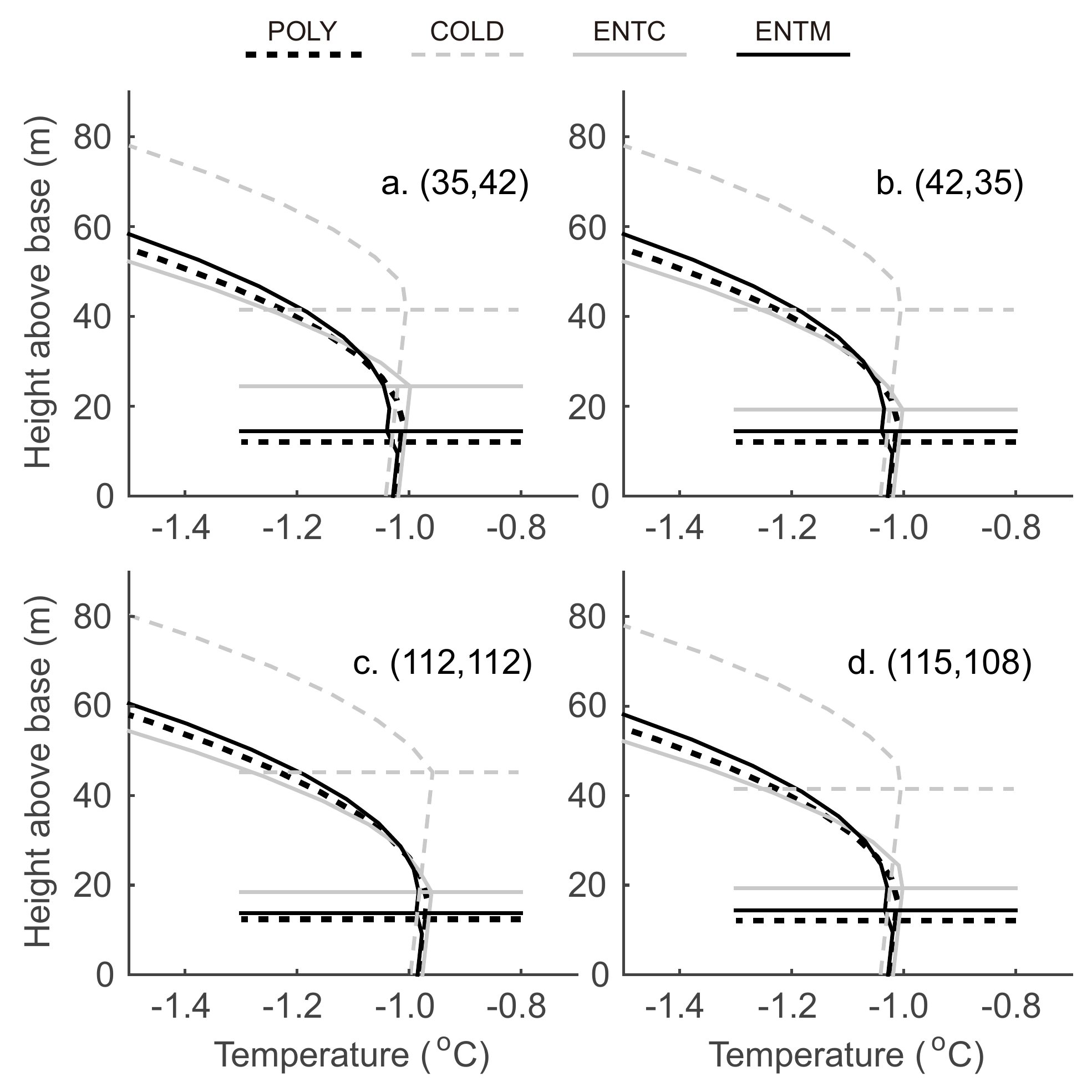}
\caption{EISMINT experiment~A1:
Vertical temperature profiles after 200\,ka model time
at four different positions near the diagonal symmetry axis
of the ice sheet
(marked by their respective grid indices $(i,j)$,
see Fig.~\ref{fig:expA_10km_20a_surftopo})
in the zone of maximum thickness of the temperate layer.
Grid resolution $\Delta{}x=10\,\mathrm{km}$,
time step $\Delta{}t=2\,\mathrm{a}$.
Black dashed: POLY scheme,
gray dashed: COLD scheme,
gray solid: ENTC scheme,
black solid: ENTM scheme
(see Sect.~\ref{sec:thermodyn_solvers} for the scheme codes).
The horizontal lines mark the CTS positions.}
\label{fig:sico_plots_profiles}
\end{center}
\end{figure}

For the same set-up, Fig.~\ref{fig:sico_plots_profiles} shows four
steady-state temperature profiles (at $t=200\,\mathrm{ka}$) in the
bottom-most 80\,m of the ice. They are all located near the margin,
$\sim\!{}520\,\mathrm{km}$ away from the center, where the largest
thicknesses of the temperate layer occur. The positions of the profiles
are indicated in Fig.~\ref{fig:expA_10km_20a_surftopo}. The profiles
show the typical patterns of the four different schemes concerning the
position of the CTS, how they match the transition conditions at the CTS
and the influence of the discrete grid. Evidently, the POLY
scheme produces the best solutions. The computed temperature
profiles are both continuous and smooth across the CTS, thus fulfilling
the transitions conditions for melting conditions. This justifies our
decision (already stated in the last paragraph of the introduction)
to use the solutions computed with the POLY scheme as a reference to assess
the performance of the other three schemes.

In the depicted four profiles, the COLD scheme overestimates the
thickness of the temperate layer by a factor $\sim\!{}3.5$.
This overestimation, which was already
reported by \citet{Greve1997b} for simulations of the Greenland ice
sheet with the SICOPOLIS model, is paralleled by a violation of the
transition condition (\ref{eq:gradT-CTS}): the temperature gradient 
across the CTS is not continuous (most pronounced in
Fig.~\ref{fig:sico_plots_profiles}c).

The profiles obtained with the ENTC scheme also
overestimate the thickness of the temperate layer, but to a lesser
extent than the cold-ice method (factor $\sim\!{}1.5$--2). Like for
the COLD scheme, the transition condition (\ref{eq:gradT-CTS})
(or its equivalent enthalpy form, Eq.~(\ref{eq:gradh-CTS}))
is not fulfilled at the CTS.

The temperate layer thicknesses found by the ENTM scheme are most
similar to those produced by the POLY scheme. Further, the temperature
profiles all show a continuous gradient across the CTS, that
is, they fulfill the transition condition (\ref{eq:gradT-CTS}) (or the
equivalent form (\ref{eq:gradh-CTS})). However, the price to pay for its
enforcement by the corrector step described in
Sect.~\ref{ssec:thermodyn_solvers_4} is that the temperature itself
shows a slight discontinuity of $\sim\!{}0.02$--0.03\,K. Ultimately,
this results from the fact that, in a one-layer scheme, the positioning
of the CTS is naturally limited by the grid resolution (while, in the
two-layer POLY scheme, arbitrary precision can be achieved).

Comparing the temperature profiles in the cold-ice layer above the CTS
shows that, despite the differences in the position of the CTS, the
results produced by the POLY, ENTM and ENTC schemes are very close to
each other. In detail, for all four profiles, the ENTM scheme is slightly
warmer, while the ENTC scheme is slightly colder than the POLY scheme.
By contrast, the COLD scheme produces notably ($\sim\!{}0.4\,\mathrm{K}$)
higher temperatures.

For the POLY, ENTM and ENTC schemes, the majority of grid points
within the temperate layer reach a water content of
$W=W_\mathrm{max}=0.01$, which is the prescribed threshold value of the
simple drainage model (see the last paragraph of
Sect.~\ref{ssec:thermodynamics_1}). Therefore, we refrain from showing
the water content in a separate figure. The COLD scheme is
an exception because it does not account for any water content; in other
words, the water content is zero everywhere.

\begin{figure}[htb]
\begin{center}
\includegraphics[width=100mm]{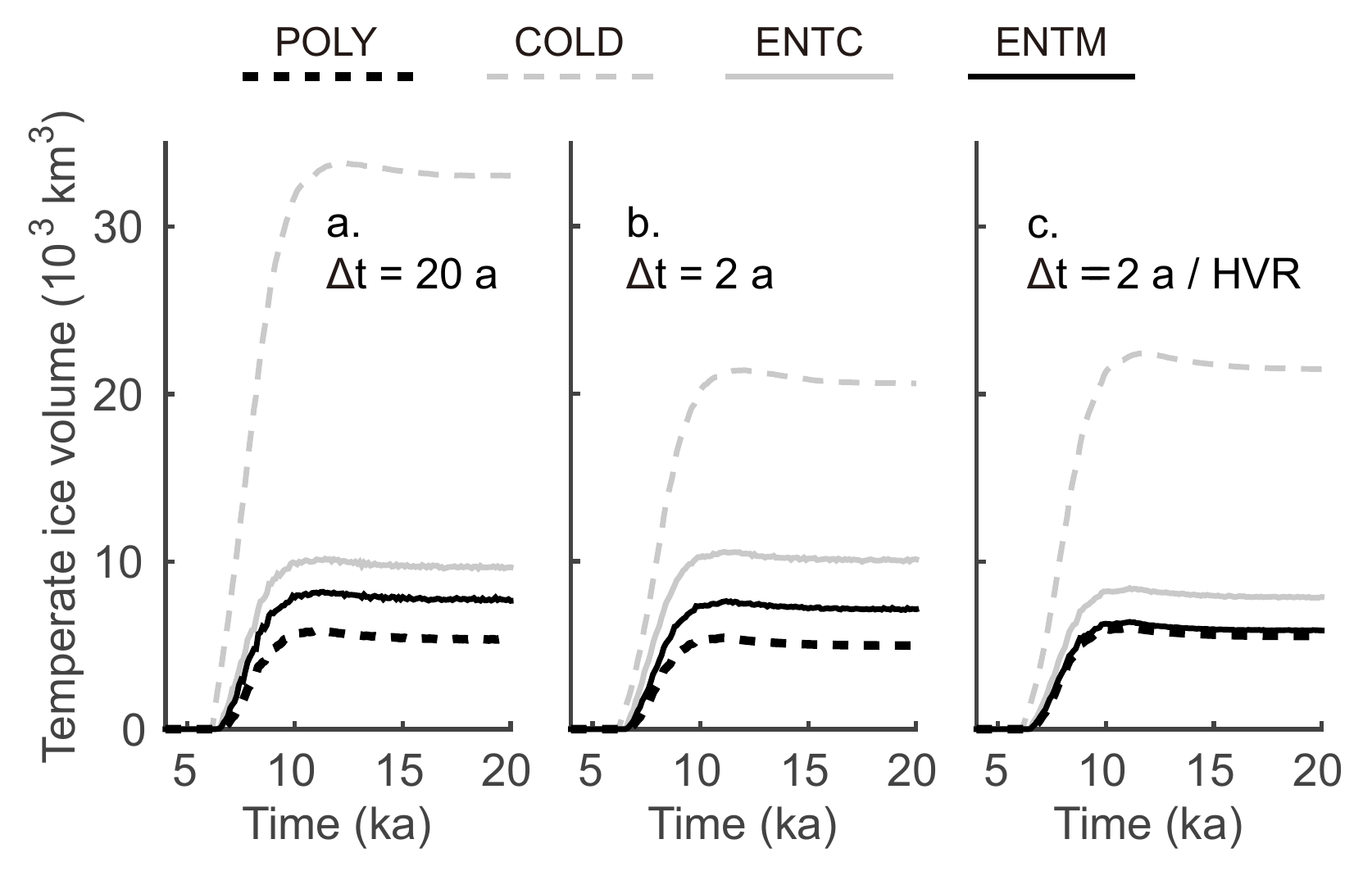}
\caption{EISMINT experiment~A1:
Evolution of the temperate ice volume over the first 20\,ka
model time.
Black dashed: POLY scheme,
gray dashed: COLD scheme,
gray solid: ENTC scheme,
black solid: ENTM scheme
(see Sect.~\ref{sec:thermodyn_solvers} for the scheme codes).
Grid resolution $\Delta{}x=10\,\mathrm{km}$.
a: Time step $\Delta{}t=20\,\mathrm{a}$.
b: Time step $\Delta{}t=2\,\mathrm{a}$.
c: Time step $\Delta{}t=2\,\mathrm{a}$, high vertical resolution
   (see Sect.~\ref{sec:setup} for details).}
\label{fig:sico_plots_series}
\end{center}
\end{figure}

In the following, we focus on the volume and thickness distribution of
the simulated temperate ice layers. This is because (a) the thermal
conditions at and near the ice base (where most of the shearing takes
place) are most relevant for ice flow, and (b) within the temperate ice,
the water content usually reaches the value $W_\mathrm{max}=0.01$ (see
above). Figure \ref{fig:sico_plots_series} shows the evolution of the
temperate ice volume for experiment~A1, run with the four thermodynamics
schemes, the grid spacing $\Delta{}x=10\,\mathrm{km}$, the two time
steps $\Delta{}t=20\,\mathrm{a}$ and $2\,\mathrm{a}$, and both the
standard and high vertical resolution (``hvr'', see Sect.~\ref{sec:setup})
set-ups. The period is limited to the first 20\,ka of the total 200\,ka
model time, during which most of the changes take place.

For all shown cases, the temperate ice volume starts forming 6--7\,ka
after the initiation of the ice-sheet build-up, goes through a weak
maximum and then converges towards a steady-state value. The main
difference between the larger ($20\,\mathrm{a}$, panel a) and the
smaller ($2\,\mathrm{a}$, panel b) time step is the behavior of the
COLD scheme. Compared to the POLY scheme, for $\Delta{}t=20\,\mathrm{a}$
it overpredicts the temperate ice volume by more than a factor 6, while
for $\Delta{}t=2\,\mathrm{a}$ the overprediction factor is $\sim\!{}4$.
The results of the POLY, ENTM and ENTC schemes do not change much; ENTC
produces $\sim\!{}2\times$ more and ENTM $\sim\!{}1.4\times$ more
temperate ice than POLY. Employing the high vertical resolution (panel
c) reduces the discrepancy between the POLY and ENTM schemes to less
than 10\%, and the overprediction by the ENTC scheme to a factor
$\sim\!{}1.4$, while there is little influence on the poor performance
of the COLD scheme.

For both time steps and the standard vertical resolution (panels a, b),
the ENTC and ENTM schemes show some high-frequency noise in the
evolution of the temperate ice volume (amplitudes less than 2\% of the
total temperate ice volume), while the results produced by the POLY
scheme and the COLD scheme are smoother. The noisiness is clearly
reduced (but still visible) for the high vertical resolution (panel c).
This makes the one-layer approach of ENTC and ENTM, with its positioning
of the CTS at a discrete grid point (uppermost grid point of the
temperate layer, see Sects.~\ref{ssec:thermodyn_solvers_3} and
\ref{ssec:thermodyn_solvers_4}) a likely cause. However, the
one-dimensional simulations carried out by \citet{Blatter&Greve2015}
for a parallel-sided slab did not show any noise, so that its
occurrence it also related to the three-dimensionality of the problem
considered here (see below).

\begin{figure}
\begin{center}
\includegraphics[width=110mm]{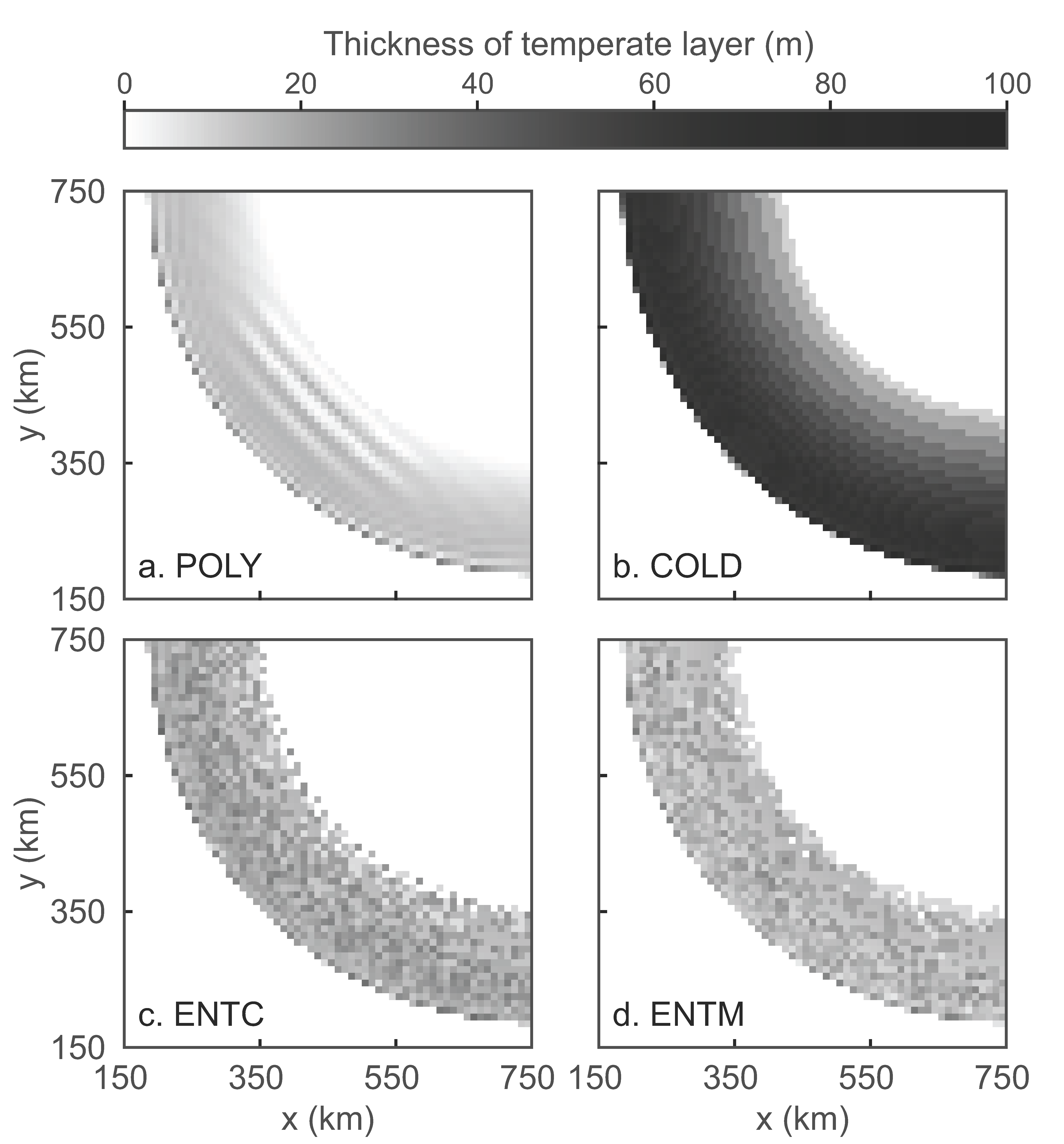}
\caption{EISMINT experiment~A1:
Thickness of the basal temperate ice layer after 200\,ka model time.
a: POLY scheme,
b: COLD scheme,
c: ENTC scheme,
d: ENTM scheme 
(see Sect.~\ref{sec:thermodyn_solvers} for the scheme codes).
Grid resolution $\Delta{}x=10\,\mathrm{km}$,
time step $\Delta{}t=20\,\mathrm{a}$.}
\label{fig:expA1_10km_20a}
\end{center}
\end{figure}

\begin{figure}
\begin{center}
\includegraphics[width=110mm]{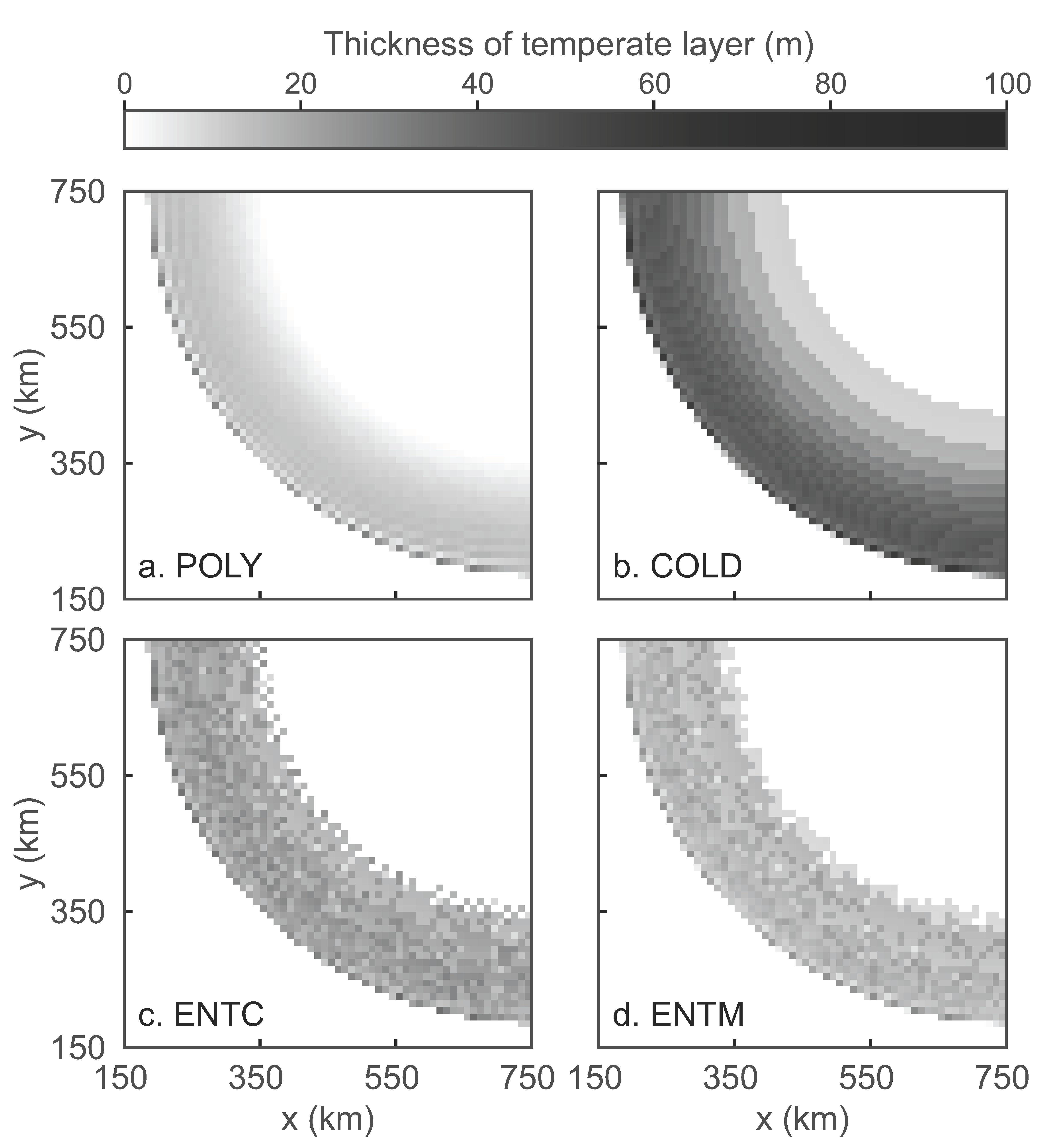}
\caption{Same as Fig.~\ref{fig:expA1_10km_20a},
but grid resolution $\Delta{}x=10\,\mathrm{km}$,
time step $\Delta{}t=2\,\mathrm{a}$.}
\label{fig:expA1_10km_02a}
\end{center}
\end{figure}

\begin{figure}
\begin{center}
\includegraphics[width=110mm]{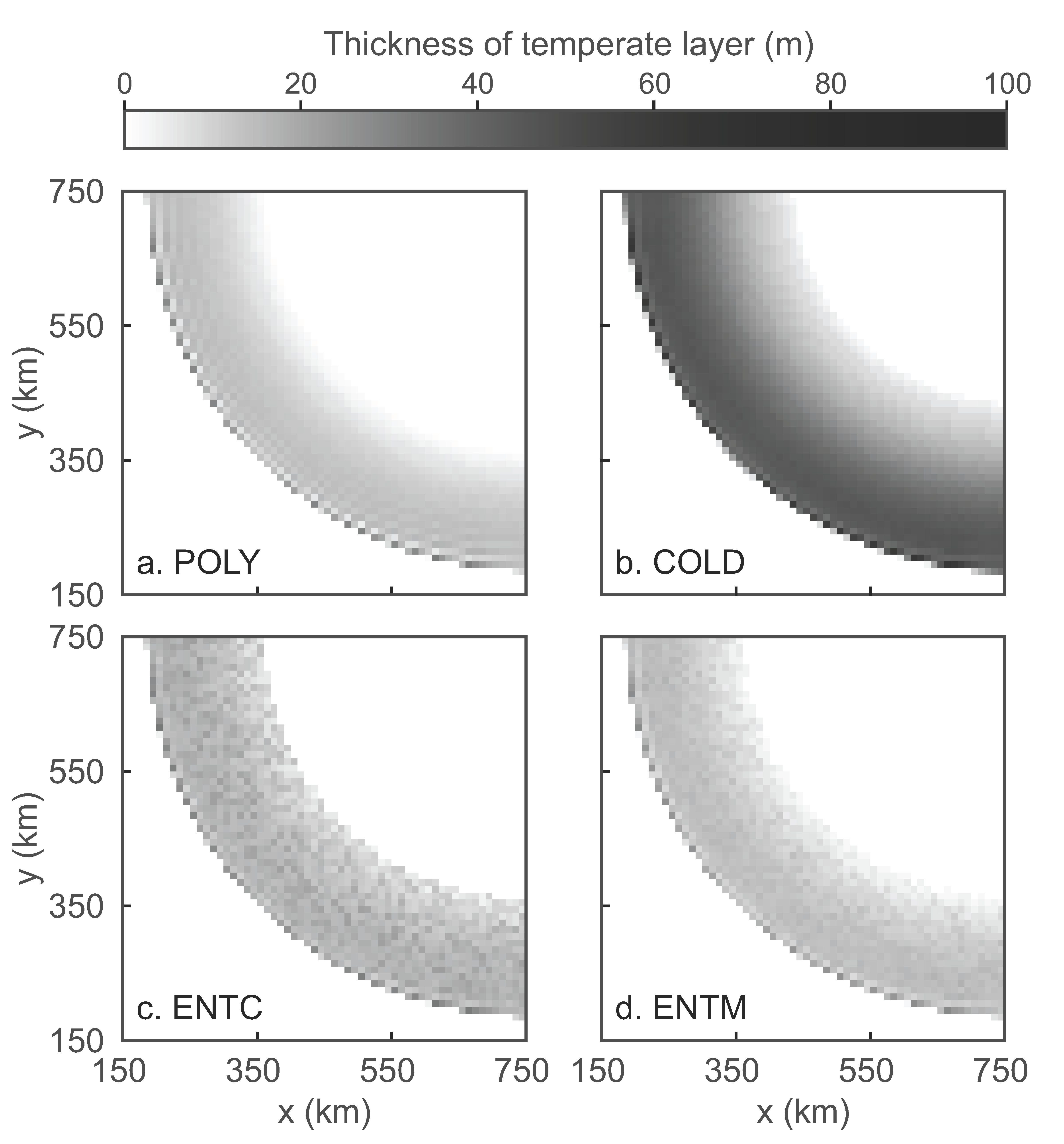}
\caption{Same as Fig.~\ref{fig:expA1_10km_20a},
but grid resolution $\Delta{}x=10\,\mathrm{km}$,
time step $\Delta{}t=2\,\mathrm{a}$,
high vertical resolution (see Sect.~\ref{sec:setup} for details).}
\label{fig:expA1_10km_02a_hvr}
\end{center}
\end{figure}

Figures \ref{fig:expA1_10km_20a}--\ref{fig:expA1_10km_02a_hvr} show maps
of the steady-state thickness of the temperate layer for experiment~A1,
all four thermodynamics solvers and the same set-ups as in
Fig.~\ref{fig:sico_plots_series}. The decreasing amount of temperate ice
in the order COLD $>$ ENTC $>$ ENTM $>$ POLY is clearly visible.
The larger time step in Fig.~\ref{fig:expA1_10km_20a} compared to
Fig.~\ref{fig:expA1_10km_02a} mainly affects the results of the POLY and
COLD schemes. For the COLD scheme, this was already discussed above.
For the POLY scheme, a numerical instability occurs in the form of a
waviness of the thickness of the temperate layer oriented diagonally to
the $x$- and $y$-axes. The smaller time step used in
Fig.~\ref{fig:expA1_10km_02a} resolves this problem.

The larger temporal noise produced by the two enthalpy
schemes (as discussed above) finds its counterpart in larger spatial
noise (compare the respective panels c and d to a and b). The two types
of noise are strongly related to each other. Both are largely unaffected
by the time step (Fig.~\ref{fig:expA1_10km_02a}
vs.\ Fig.~\ref{fig:expA1_10km_20a}), but significantly reduced by the
higher vertical resolution (Fig.~\ref{fig:expA1_10km_02a_hvr}
vs.\ Fig.~\ref{fig:expA1_10km_02a}). For the
small-time-step/high-vertical-resolution set-up used in
Fig.~\ref{fig:expA1_10km_02a_hvr}, the results of the POLY and ENTM
schemes are very similar to each other.

Further, in all cases, the thickness of the temperate layer tends to
increase gradually with distance from the center of the ice sheet, and
then decreases sharply near the ice margin. This confirms for our new
simulations what was already claimed in Sect.~\ref{sec:intro}, and it
is the reason for the fact that freezing conditions at the CTS can
hardly be resolved, thus justifying that we ignored them for this study.
Owing to the incompatible symmetries of the ice sheet geometry
(circularly symmetric) and the numerical grid (aligned to the $x$- and
$y$-axes), the thin rings of largest temperate layer thicknesses near
the margin are in all cases somewhat discontinuous.

\begin{table}
  \centering
  \begin{tabular}{lcccc} \hline
  Exp.~A1 set-up & Volume & Area & Temp.\,volume & Melt \\
  (scheme, $\Delta{}x$, $\Delta{}t$)
  & ($10^6\,\mathrm{km}^3$) & ($10^6\,\mathrm{km}^2$)
  & ($10^4\,\mathrm{km}^3$) & fraction \\ \hline
  POLY, 10\,km, 20\,a & 2.105 & 1.045 & 0.537 & 0.675
  \\
  COLD, 10\,km, 20\,a & 2.109 & 1.043 & 3.319 & 0.671
  \\
  ENTC, 10\,km, 20\,a & 2.097 & 1.044 & 0.968 & 0.673
  \\
  ENTM, 10\,km, 20\,a & 2.113 & 1.046 & 0.775 & 0.674
  \\ \hline
  POLY, 10\,km,  2\,a & 2.109 & 1.046 & 0.501 & 0.675
  \\
  COLD, 10\,km,  2\,a & 2.120 & 1.044 & 2.076 & 0.675
  \\
  ENTC, 10\,km,  2\,a & 2.096 & 1.044 & 1.008 & 0.676
  \\
  ENTM, 10\,km,  2\,a & 2.110 & 1.046 & 0.718 & 0.674
  \\ \hline
  POLY,  5\,km,  2\,a & 2.123 & 1.050 & 0.466 & 0.675
  \\
  COLD,  5\,km,  2\,a & 2.133 & 1.049 & 2.009 & 0.674
  \\
  ENTC,  5\,km,  2\,a & 2.110 & 1.050 & 0.977 & 0.675
  \\
  ENTM,  5\,km,  2\,a & 2.124 & 1.050 & 0.689 & 0.675
  \\ \hline
  POLY, 10\,km,  2\,a (hvr) & 2.105 & 1.045 & 0.559 & 0.695
  \\
  COLD, 10\,km,  2\,a (hvr) & 2.120 & 1.044 & 2.159 & 0.696
  \\
  ENTC, 10\,km,  2\,a (hvr) & 2.101 & 1.044 & 0.790 & 0.694
  \\
  ENTM, 10\,km,  2\,a (hvr) & 2.105 & 1.045 & 0.595 & 0.694
  \\ \hline
  \end{tabular}
  \caption{EISMINT experiment~A1:
  Mean values between $t=190\,\mathrm{ka}$ and 200\,ka for
  the ice volume, the ice area, the volume of temperate ice and the
  melt fraction (fraction of basal ice at the pressure melting point).
  See Sect.~\ref{sec:thermodyn_solvers} for the scheme codes.
  The abbreviation ``hvr'' means ``high vertical resolution''
  (see Sect.~\ref{sec:setup} for details).}
  \label{tab_results_expA1}
\end{table}

\begin{table}
  \centering
  \begin{tabular}{lc} \hline
  Exp.~A1 set-up & Computing \\
  (scheme, $\Delta{}x$, $\Delta{}t$) & time (hrs) \\ \hline
  POLY, 10\,km,  2\,a & 7.7
  \\
  COLD, 10\,km,  2\,a & 4.2
  \\
  ENTC, 10\,km,  2\,a & 7.0
  \\
  ENTM, 10\,km,  2\,a & 8.2
  \\ \hline
  \end{tabular}
  \caption{Computing times for EISMINT experiment~A1 with the reference
  set-up ($\Delta{}x=10\,\mathrm{km}$, $\Delta{}t=2\,\mathrm{a}$),
  run with the Intel Fortran Compiler 15.0.3 for Linux
  (optimization options -xHOST -O3 -no-prec-div)
  on a 12-Core Intel Xeon E5-2697~v2 (2.7~GHz) PC under openSUSE 13.1 (64~bit).
  See Sect.~\ref{sec:thermodyn_solvers} for the scheme codes.}
  \label{tab_cpu_times_expA1}
\end{table}

A synopsis of the main results of the 16 set-ups of experiment~A1
(all set-ups shown in
Figs.~\ref{fig:expA1_10km_20a}--\ref{fig:expA1_10km_02a_hvr}, plus the
high-horizontal-resolution case
$\Delta{}x=5\,\mathrm{km}$, $\Delta{}t=2\,\mathrm{a}$) is
given in Table~\ref{tab_results_expA1}. The volume of temperate ice is
in the range of $\sim\!{}0.25$--1\% of the total ice volume and, as
discussed above, varies strongly across the different thermodynamics
schemes. By contrast, the melt fraction (area ratio of basal temperate
ice to total ice) is much larger ($\sim\!{}2/3$) and varies very little
across all 16 set-ups. For the total ice volume and area, the strongest
systematic (but still small) influence arises from the horizontal
resolution: On average, $\Delta{}x=5\,\mathrm{km}$ produces an ice
sheet with 0.7\% more volume and 0.5\% more area than
$\Delta{}x=10\,\mathrm{km}$. By contrast, the influences of the
thermodynamics solver, the time step and the vertical resolution are
unsystematic and even smaller.

In Table~\ref{tab_cpu_times_expA1}, the computing times of the four
thermodynamics schemes are compared for Experiment A1 with
$\Delta{}x=10\,\mathrm{km}$ and $\Delta{}t=2\,\mathrm{a}$. The
differences between the schemes are significant. As expected, the most
simple, but physically inadequate COLD scheme is by far the fastest,
with a gain of more than 40\% compared to the POLY scheme. The
differences between the POLY, ENTM and ENTC schemes are smaller. ENTC is
about 10\% faster than POLY, while ENTM is about 6\% slower. The reason
for the relative slowness of the enthalpy schemes, despite greater
simplicity compared to the POLY scheme, is that, in the SICOPOLIS
implementation, in each time step the computed enthalpy field is
converted back to temperature and water content. This must be done
three-dimensionally and is thus costly. Codes that are designed
as enthalpy-based from the outset (whereas the thermodynamics of
SICOPOLIS is natively based on temperature and water content)
can probably avoid these conversions, which opens the possibility
for greater efficiency of the enthalpy schemes.

\begin{figure}
\begin{center}
\includegraphics[width=110mm]{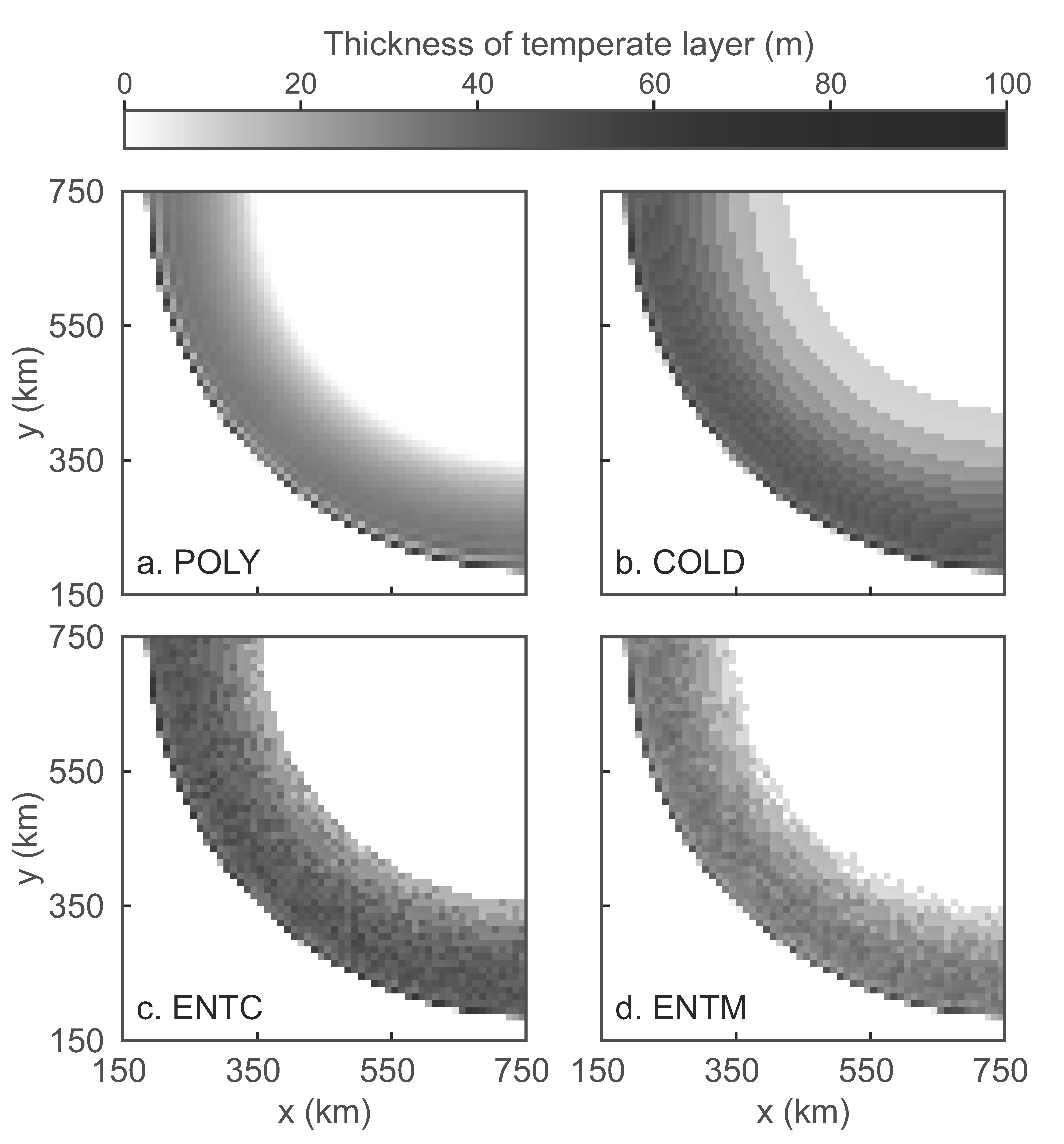}
\caption{Same as Fig.~\ref{fig:expA1_10km_20a},
but for EISMINT experiment A2,
grid resolution $\Delta{}x=10\,\mathrm{km}$,
time step $\Delta{}t=2\,\mathrm{a}$.}
\label{fig:expA2_10km_02a}
\end{center}
\end{figure}

\begin{figure}
\begin{center}
\includegraphics[width=110mm]{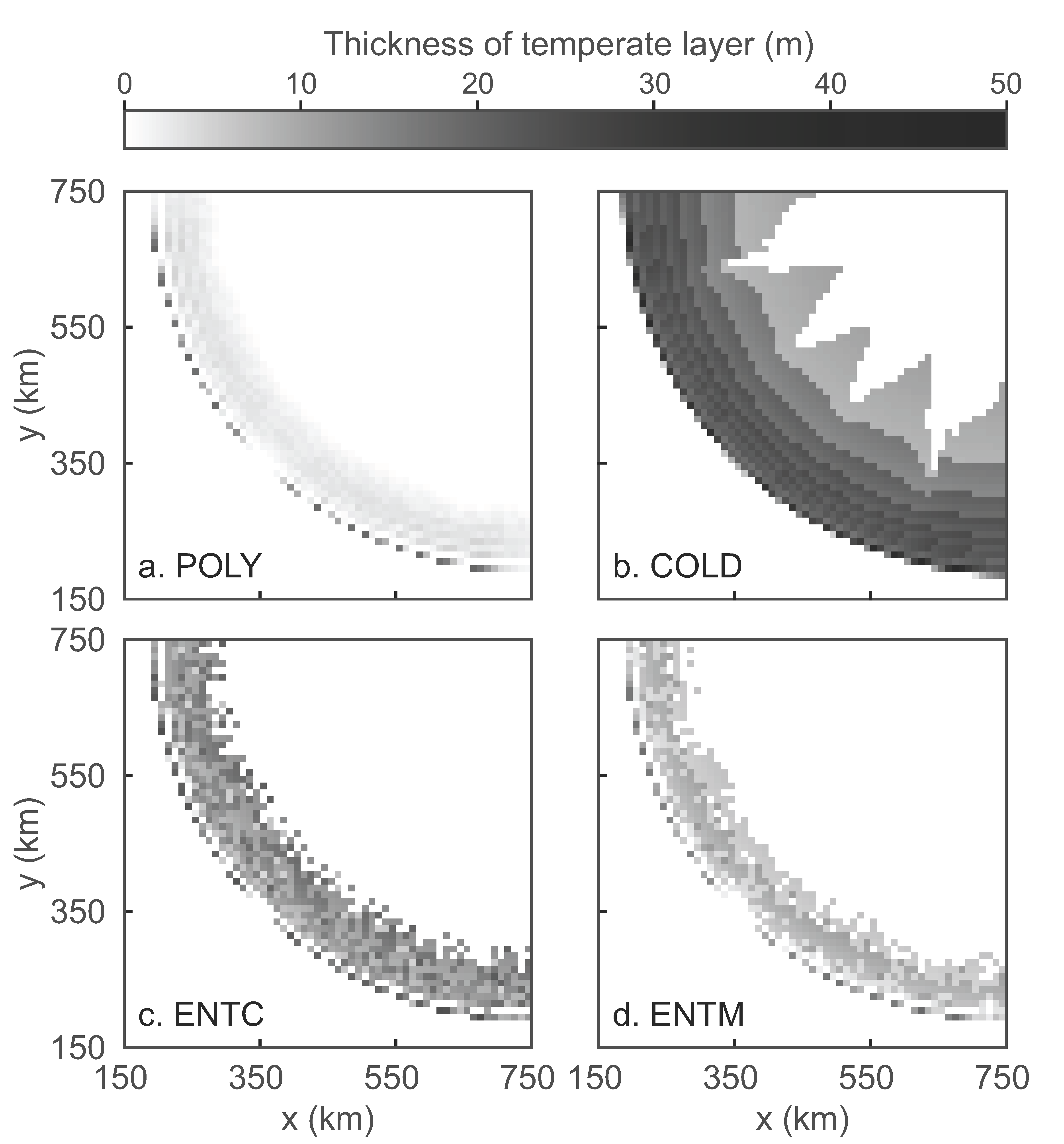}
\caption{Same as Fig.~\ref{fig:expA1_10km_20a},
but for EISMINT experiment H1,
grid resolution $\Delta{}x=10\,\mathrm{km}$,
time step $\Delta{}t=2\,\mathrm{a}$.}
\label{fig:expH1_10km_02a}
\end{center}
\end{figure}

For experiment A2, the water-content-dependent rate factor for temperate
ice has been replaced by a constant rate factor like in the original
EISMINT set-up (Sect.~\ref{ssec:setup2}). The resulting steady-state
thickness of the temperate layer for all four thermodynamics solvers and
the combination $\Delta{}x=10\,\mathrm{km}$, $\Delta{}t=2\,\mathrm{a}$
is shown in Fig.~\ref{fig:expA2_10km_02a}, and an overview of the main
results is given in Table~\ref{tab_results_expA1A2H1}. The COLD scheme
is unaffected by the difference between experiments A1 and A2 because it
does not account for any water content, and consequently the results are
the same. By contrast, the POLY, ENTM and ENTC schemes produce roughly
two times thicker temperate ice layers for experiment A2. The reason is
that the simulated temperate ice is stiffer, which reduces the advective
transport of cold ice towards the base, so that the growth of the
temperate ice layers is favored compared to experiment A1. It is
interesting to note that, for experiment A2, the temperate ice volumes 
computed by the POLY and ENTM schemes agree very well (within 3\%) for
the standard vertical resolution employed here. For experiment A1, 
a similarly close agreement is only achieved with the high vertical
resolution (see discussion above).

There is no notable influence of experiment A2 on the melt fraction. The
total ice area is essentially unaffected as well, while the total ice
volume is systematically slightly (on average $\sim\!{}0.8\%$) larger
than in experiment A1.

\begin{table}
  \centering
  \begin{tabular}{lcccc} \hline
  Exp.~A1 set-up & Volume & Area & Temp.\,volume & Melt \\
  (scheme, $\Delta{}x$, $\Delta{}t$)
  & ($10^6\,\mathrm{km}^3$) & ($10^6\,\mathrm{km}^2$)
  & ($10^4\,\mathrm{km}^3$) & fraction \\ \hline
  POLY, 10\,km,  2\,a & 2.109 & 1.046 & 0.501 & 0.675
  \\
  COLD, 10\,km,  2\,a & 2.120 & 1.044 & 2.076 & 0.675
  \\
  ENTC, 10\,km,  2\,a & 2.096 & 1.044 & 1.008 & 0.676
  \\
  ENTM, 10\,km,  2\,a & 2.110 & 1.046 & 0.718 & 0.674
  \\ \hline
  Exp.~A2 set-up & Volume & Area & Temp.\,volume & Melt \\
  (scheme, $\Delta{}x$, $\Delta{}t$)
  & ($10^6\,\mathrm{km}^3$) & ($10^6\,\mathrm{km}^2$)
  & ($10^4\,\mathrm{km}^3$) & fraction \\ \hline
  POLY, 10\,km,  2\,a & 2.124 & 1.044 & 1.245 & 0.675
  \\
  COLD, 10\,km,  2\,a & 2.120 & 1.044 & 2.076 & 0.675
  \\
  ENTC, 10\,km,  2\,a & 2.120 & 1.044 & 1.956 & 0.675
  \\
  ENTM, 10\,km,  2\,a & 2.123 & 1.044 & 1.273 & 0.676
  \\ \hline
  Exp.~H1 set-up & Volume & Area & Temp.\,volume & Melt \\
  (scheme, $\Delta{}x$, $\Delta{}t$)
  & ($10^6\,\mathrm{km}^3$) & ($10^6\,\mathrm{km}^2$)
  & ($10^4\,\mathrm{km}^3$) & fraction \\ \hline
  POLY, 10\,km,  2\,a & 2.094 & 1.045 & 0.079 & 0.668
  \\
  COLD, 10\,km,  2\,a & 2.090 & 1.041 & 1.149 & 0.659
  \\
  ENTC, 10\,km,  2\,a & 2.087 & 1.045 & 0.365 & 0.664
  \\
  ENTM, 10\,km,  2\,a & 2.096 & 1.045 & 0.176 & 0.664
  \\ \hline
  \end{tabular}
  \caption{EISMINT experiments~A1, A2 and H1:
  Mean values between $t=190\,\mathrm{ka}$ and 200\,ka for
  the ice volume, the ice area, the volume of temperate ice and the
  melt fraction (fraction of basal ice at the pressure melting point).
  See Sect.~\ref{sec:thermodyn_solvers} for the scheme codes.}
  \label{tab_results_expA1A2H1}
\end{table}

We discussed above that, for experiment A1, the POLY scheme with the
large-time-step setting $\Delta{}x=10\,\mathrm{km}$,
$\Delta{}t=20\,\mathrm{a}$ produces a temperate-layer thickness that
shows a waviness oriented diagonally to the $x$- and $y$-axes
(Fig.~\ref{fig:expA1_10km_20a}a). For experiment A2 and
$\Delta{}x=10\,\mathrm{km}$, $\Delta{}t=20\,\mathrm{a}$, this
instability does not occur, and the thickness of the temperate
layer (not shown) is very similar to that obtained for the setting
$\Delta{}x=10\,\mathrm{km}$, $\Delta{}t=2\,\mathrm{a}$ shown in
Fig.~\ref{fig:expA2_10km_02a}a. Hence, the instability is linked
to the enhanced fluidity in the temperate layer due to the
water-content-dependent rate factor employed in experiment A1.

For experiment~H1, basal sliding over temperate- and
near-temperate-based areas has been added to the set-up of experiment~A1
(Sect.~\ref{ssec:setup3}). Like for experiment A2, we only discuss the
combination $\Delta{}x=10\,\mathrm{km}$, $\Delta{}t=2\,\mathrm{a}$.
Figure~\ref{fig:expH1_10km_02a} depicts the steady-state thickness of
the temperate layer for the four thermodynamics solvers, and
Table~\ref{tab_results_expA1A2H1} gives an overview of the results.
Comparing Figs.~\ref{fig:expA1_10km_02a} and \ref{fig:expH1_10km_02a} as
well as the entries in Table~\ref{tab_results_expA1A2H1} reveals that
the addition of some basal sliding reduces the volume of temperate ice
strongly. This is a consequence of increased advection of cold surface
ice downward and outward, and of reduced strain heating. Like for
experiments~A1 and A2, the computed temperate ice volume decreases in
the order COLD $>$ ENTC $>$ ENTM $>$ POLY. However, even the ENTM scheme
overpredicts the temperate ice volume by more than a factor 2 compared
to the POLY scheme. This is because the thicknesses are so small that
any one-layer scheme does not resolve them well (unless a very high
vertical resolution is employed), whereas the performance of the
two-layer POLY scheme is not limited by this problem.

Figure \ref{fig:expH1_10km_02a}b shows further that in the distribution
of temperate ice obtained with the COLD scheme some spokes emerge. The
spokes are also present in similar form in the results of the three
other schemes when the basal temperature is plotted instead of the
thickness of the temperate ice layer (not shown). This phenomenon was
already observed in the original article on the EISMINT Phase~2
Simplified Geometry Experiments \citep{Eismint2000}. It is most likely a
consequence of the combined effects of a thermoviscous flow instability
and the symmetry of the numerical grid \citep{Bueler&etal2007}.

Since the sliding coefficient in experiment~H1 is small,
the melt fraction and the total ice volume
only decrease slightly compared to those of experiment~A1, and the
ice area is virtually unchanged. This underlines the very high
sensitivity of the temperate ice layer to changed flow conditions.
For the original experiment~H with a 10 times larger sliding
coefficient (results not shown), the large-scale response of the ice
sheet is more pronounced. For this scenario, only the COLD scheme
produces a temperate layer of non-zero thickness, while the other schemes
merely produce a temperate base.

\section{Discussion and conclusion}
\label{sec:discussion}

We tested four different thermodynamics solvers in the ice sheet model
SICOPOLIS. Two of them are the previously existing polythermal two-layer
(POLY) and cold-ice (COLD) schemes, while the other two are the newly
implemented conventional one-layer enthalpy (ENTC) and melting-CTS
one-layer enthalpy (ENTM) schemes. The ENTC scheme goes
back to the study by \citet{AschwandenBueler2012}, while the ENTM
scheme was introduced by \citet{Blatter&Greve2015} for a
one-dimensional ice slab. Extension to the full, three-dimensional
problem was straightforward because, due to the neglect of horizontal
diffusive heat fluxes, the computation of the thermodynamic fields
(temperature and water content or enthalpy) is essentially
one-dimensional for each column. Horizontal advection merely plays the
role of an additional source term for the vertical profiles and does not
constitute a problem for the implementation of the scheme.

We used two scenarios from the suite of EISMINT (European Ice Sheet
Modeling INiTiative) Phase~2 Simplified Geometry Experiments
\citep{Eismint2000} as our model problems, one without and one with
basal sliding. Like in the one-dimensional study carried out by
\citet{Blatter&Greve2015}, the results computed with the POLY
scheme were assessed to be most reliable and thus chosen as a
reference.

As it was already reported by \citet{Greve1997b}, the COLD scheme
strongly overpredicts thicknesses of temperate ice layers, and thus
temperate ice volumes. Choosing a small time step reduces the amount of
overprediction to a certain extent, but the performance of the COLD
scheme (which is not energy-conserving and thus physically incorrect
anyway) is not satisfactory.

The studies by \citet{Kleiner&al2015} and \citet{Blatter&Greve2015},
both carried out for a Canadian-type parallel-sided slab, found that the
ENTC scheme, even though it does not enforce the transition conditions
at the CTS explicitly, still fulfills them for the case of melting
conditions (continuity of the enthalpy and the sensible heat flux),
provided that the discontinuity of the enthalpy diffusivity
(\ref{eq:enthalpy-diffusivity}) at the CTS is properly accounted for in
the discretization of the diffusion term in the enthalpy equation
(\ref{eq:enthalpy}). This could not be confirmed in the present study:
even though the ENTC scheme has been implemented in SICOPOLIS in the
same way as described by \citet{Blatter&Greve2015}, the scheme fails to
produce a continuous temperature gradient across the CTS and
overpredicts temperate ice layer thicknesses, albeit to a smaller
degree than the COLD scheme (always compared to the references provided
by the POLY scheme). We speculate that this is a consequence of the
three-dimensionality of the problem investigated here, most likely due
to the additional horizontal advection terms in the enthalpy equation.

In contrast to the ENTC scheme, the continuity of the temperature
gradient across the CTS is explicitly enforced by the ENTM scheme. This
leads to a better match of computed temperate ice layer thicknesses to
those computed by the POLY scheme; however, some overprediction remains.
The price to pay for enforcing a continuous temperature gradient across
the CTS, while the positioning of the CTS is limited by the grid
resolution, is a slight discontinuity of the ice temperature itself.

When comparing the temperature profiles in the ice column above
the CTS, the ENTC and ENTM schemes both perform well compared to the
POLY scheme, while the temperatures produced by the COLD scheme are
too high. The water content in the temperate ice layer below the CTS
usually reaches the threshold value of 0.01 (1\%) for the POLY, ENTC
and ENTM schemes, while the COLD scheme does not compute any water
content.

For large time steps in conjunction with a water-content-dependent rate
factor, the POLY scheme can produce a wavy distribution of the
temperate-layer thickness. Therefore, some care is required to choose a
sufficiently small time step to avoid this instability. However, in
contrast to the ``academic'' EISMINT application considered here,
real-world applications require small time steps for overall numerical
stability anyway. At $10\,\mathrm{km}$ horizontal resolution, Greenland
simulations with SICOPOLIS typically require time steps of
$\sim{}1\,\mathrm{a}$, and Antarctica simulations with coupled
sheet--shelf dynamics even smaller ones. So this instability is most
likely not a matter of concern for most practical applications with
complex topographies.

The ENTC and ENTM schemes produce some temporal and spatial noise
for the volume and thickness of the temperate ice. This does not
happen for the POLY and COLD schemes; it is related to the 
one-layer approach in which we position the CTS at the uppermost
grid point in the temperate layer, whereas the POLY scheme allows,
in principle, tracking the position of the CTS to arbitrary accuracy.
A possible improvement of the one-layer enthalpy schemes would be to
implement a sub-grid tracking of the CTS position, which would likely
allow to fulfill the continuity of the temperate and of the temperature
gradient across the CTS simultaneously and reduce the noise. However,
such a sub-grid tracking scheme will inevitably increase the
computational cost.

To sum up, for a polythermal ice sheet model, the one-layer enthalpy
schemes ENTC and ENTM are viable, easier to implement alternatives to
the POLY scheme with its slightly awkward need to handle two different
numerical domains for cold and temperate ice. ENTM is more precise than
ENTC for determining the position of the CTS, and thus the volume and
thickness of the temperate ice layer. The performance of both schemes is
good for the temperature profiles in the cold ice column above the
CTS and also the water content in the temperate ice layer below the CTS.
The computing times of all three schemes are comparable.

\section*{Acknowledgments}

We thank the two anonymous reviewers and the scientific editor,
T.\ Kameda, for their comments that helped considerably to improve the
manu\-script.
R.G.\ was supported by the Japanese Ministry of Education, Culture,
Sports, Science and Technology (MEXT) through the Green Network
of Excellence (GRENE) Arctic Climate Change Research project
and the Arctic Challenge for Sustainability (ArCS) project.
H.B.\ was supported by an Invitation Fellowship for Research in Japan
(No.\ L13525) of the Japan Society for the Promotion of Science
(JSPS).


\end{document}